\documentclass[prl,aps,twocolumn,superscriptaddress,notitlepage]{revtex4}

\usepackage{graphicx}
\usepackage{amsmath}
\usepackage{amssymb}
\usepackage{times}
\usepackage{color}
\usepackage{soul}
\usepackage{bbm}
\usepackage{bm}

\begin{document}

\title{Quantum interferometry with three-dimensional geometry}

\author{Nicol\`o Spagnolo}
\affiliation{Dipartimento di Fisica, Sapienza Universit\`{a} di Roma, Piazzale Aldo Moro 5, I-00185 Roma, Italy}
\author{Lorenzo Aparo}
\affiliation{Dipartimento di Fisica, Sapienza Universit\`{a} di Roma, Piazzale Aldo Moro 5, I-00185 Roma, Italy}
\author{Chiara Vitelli}
\affiliation{Center of Life NanoScience @ La Sapienza, Istituto Italiano di Tecnologia, Viale Regina Elena, 255, I-00185 Roma, Italy}
\affiliation{Dipartimento di Fisica, Sapienza Universit\`{a} di Roma, Piazzale Aldo Moro 5, I-00185 Roma, Italy}
\author{Andrea Crespi}
\affiliation{Istituto di Fotonica e Nanotecnologie, Consiglio Nazionale delle Ricerche (IFN-CNR), Piazza Leonardo da Vinci 32, I-20133 Milano, Italy}
\affiliation{Dipartimento di Fisica, Politecnico di Milano, Piazza Leonardo da Vinci 32, I-20133 Milano, Italy}
\author{Roberta Ramponi}
\affiliation{Istituto di Fotonica e Nanotecnologie, Consiglio Nazionale delle Ricerche (IFN-CNR), Piazza Leonardo da Vinci 32, I-20133 Milano, Italy}
\affiliation{Dipartimento di Fisica, Politecnico di Milano, Piazza Leonardo da Vinci 32, I-20133 Milano, Italy}
\author{Roberto Osellame}
\affiliation{Istituto di Fotonica e Nanotecnologie, Consiglio Nazionale delle Ricerche (IFN-CNR), Piazza Leonardo da Vinci 32, I-20133 Milano, Italy}
\affiliation{Dipartimento di Fisica, Politecnico di Milano, Piazza Leonardo da Vinci 32, I-20133 Milano, Italy}
\author{Paolo Mataloni}
\affiliation{Dipartimento di Fisica, Sapienza Universit\`{a} di Roma, Piazzale Aldo Moro 5, I-00185 Roma, Italy}
\affiliation{Istituto Nazionale di Ottica (INO-CNR), Largo E. Fermi 6, I-50125 Firenze, Italy}
\author{Fabio Sciarrino}
\affiliation{Dipartimento di Fisica, Sapienza Universit\`{a} di Roma, Piazzale Aldo Moro 5, I-00185 Roma, Italy}
\affiliation{Istituto Nazionale di Ottica (INO-CNR), Largo E. Fermi 6, I-50125 Firenze, Italy}

\begin{abstract}
\end{abstract}

\maketitle

\textbf{Quantum interferometry uses quantum resources to improve phase estimation with respect to classical methods. Here we propose and theoretically investigate a new quantum interferometric scheme based on three-dimensional waveguide devices. These can be implemented by femtosecond laser waveguide writing, recently adopted for quantum applications. In particular, multiarm interferometers include ``tritter'' and ``quarter'' as basic elements, corresponding to the generalization of a beam splitter to a 3- and 4-port splitter, respectively. By injecting Fock states in the input ports of such interferometers, fringe patterns characterized by nonclassical visibilities are expected. This enables outperforming the quantum Fisher information obtained with classical fields in phase estimation. We also discuss the possibility of achieving the simultaneous estimation of more than one optical phase. This approach is expected to open new perspectives to quantum enhanced sensing and metrology performed in integrated photonic.}

\section{Introduction}

Quantum metrology is one of the most fascinating frontiers of the science of measurement: the counterintuitive laws of quantum mechanics are exploited to maximize the amount of information extracted from an unknown sample, beating the limits imposed by classical physics. The low decoherence of photons, enabling the observation of quantum effects in an easier way, makes optical interferometry a promising candidate for demonstrating quantum enhanced sensitivity. The estimation of an optical phase $\phi$ through interferometric experiments is indeed an ubiquitous technique in physics, ranging from the investigation of fragile biological samples, such as tissues \cite{Nasr09} or blood proteins in aqueous buffer solution \cite{Cres12}, to gravitational wave measurements \cite{Goda08,Abad11}. Whereas optical interferometry relying on classical interference is intrinsically a single-particle process, quantum advantages arise when quantum-correlated states of more than one particle are employed \cite{Giov04b}, such as Greenberger-Horne-Zeilinger (GHZ) \cite{Bouw99} and NOON \cite{Boto00} states.  NOON states, in particular, allow to saturate the Heisenberg limit of sensitivity: the ultimate limit to the precision of a measurement imposed by the laws of physics.  

%
%
\begin{figure}[b!]
\centering
\includegraphics[width=0.5\textwidth]{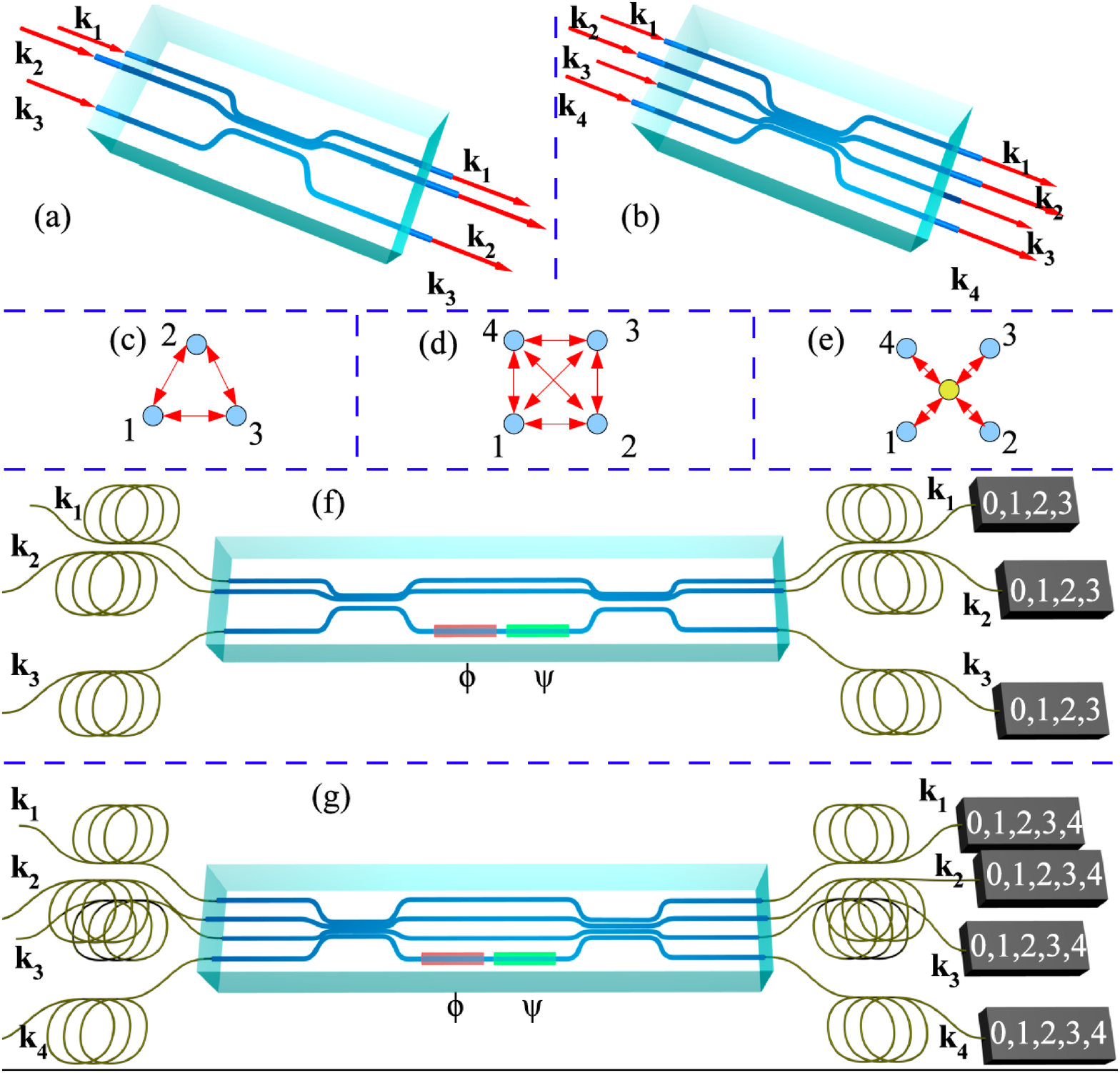}
\caption{(a) 3D structure of tritter. (b) 3-dimensional structure of quarter. (c) Geometry showing the direct coupling between the three modes of the tritter. (d) Geometry showing the direct coupling between the four modes of the quarter. (e) Indirect coupling between the four modes of the quarter by means of one ancillary mode. (f) 3-mode interferometer built by using two cascaded tritters. (g) 4-mode interferometer built by cascading two quarter devices. (f)-(g) $\phi$: phase to be measured. $\psi$: additional phase for adaptive phase estimation.}
\label{fig:fig1}
\end{figure}

Generally, the quantum advantage over classical approaches increases with the number of particles involved in the correlated probe state \cite{Giov04b}. Two-photon NOON states can be produced deterministically simply by quantum interference of two identical photons on a balanced two-mode beam-splitter. However, two-mode NOON states with more than two particles are difficult to produce. Increasing the number of modes to more than two represents an interesting possibility to extend the concept of multiparticle interferometry, as pointed out by Greenberger et al. \cite{Gree00}. This requires multi-port devices, instead of simple two-mode beam-splitters, to build multi-arm interferometers. Multi-port beam splitters can be potentially realized by properly combining several balanced two-port beam-splitters and phase shifters \cite{Zuko97,Camp00}: such implementation has anyway tight requirements on interferometric stability, making its effective realization challenging with bulk optics. In fact, the few experimental realizations of such devices are reported on fiber-based \cite{Weih96} or integrated-optics \cite{Peru11} multi-mode devices; in both cases the characterization was performed with only two-photon states. Multi-photon interferometry is, indeed, still a widely unexplored field.

Times are mature to demonstrate multi-photon and multi-port devices. In fact, on the one hand in the last few years the efficiency of quantum multi-photon sources have dramatically improved leading to several experiments with up to 8 photons \cite{Yao12}. On the other hand, the advent of integrated quantum photonics have opened exciting perspectives for the realization of scalable, miniaturized and intrinsically stable optical setups \cite{Poli08}. In particular, the ultrafast laser-writing technique \cite{Gatt08, Dell09,Osel11,Nolte03} has proved to be a powerful tool for demonstrating new quantum integrated-optics devices, able to perform quantum logic operations \cite{Cres11} as well as two-photon quantum walks \cite{Owen11,Sans12}. This technique exploits nonlinear absorption of focused femtosecond laser pulses to induce permanent and localized increase of the refractive index in transparent materials. Waveguides are directly fabricated in the material bulk by translating the sample at constant velocity along the desired path with unique three-dimensional (3D) capabilities.

In the field of quantum metrology, recent results have demonstrated the possibility of using two-mode path-entangled states in integrated structures for phase estimation below the Standard Quantum Limit (SQL) \cite{Smit09,Matt11,Shad11,Cres12}. Combination of all the above elements would enable stepping into multi-photon/multi-port quantum metrology. However, the potentials of this approach have not yet been investigated theoretically.

In this work we introduce the concept of 3D multi-photon interferometry. First, we propose novel geometries for integrated multi-arm interferometers based on three-port (tritter) and four-port (quarter) devices. Second, we theoretically study possible measurement protocols, based on the injection of multi-photon Fock states in this kind of multi-port devices, demonstrating relevant metrological advantages in phase-estimation tasks. Our results are not merely speculative since both the realization of such integrated multi-port devices \cite{Kowa05} and the generation of multi-photon quantum states \cite{Metc12} appear to be within reach of present state-of-the-art technology. 

%
%
\begin{figure*}[ht!]
\includegraphics[width=0.94\textwidth]{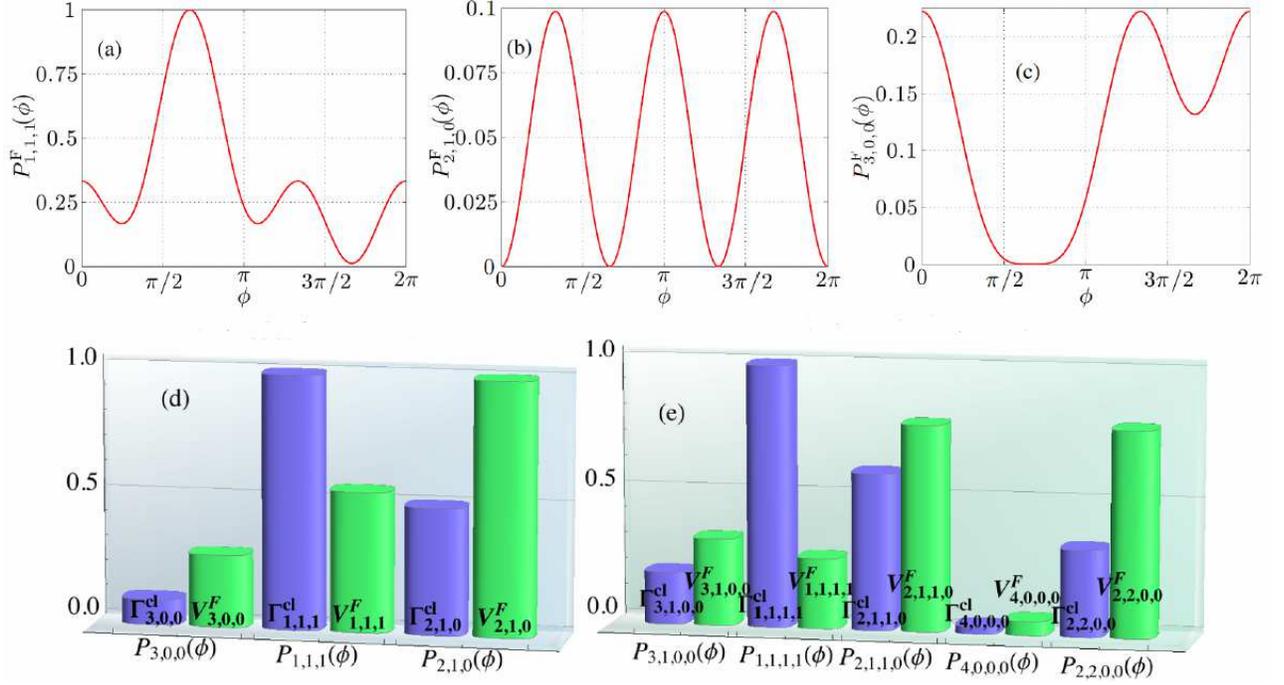}
\caption{(a)-(c) Output fringe patterns $P^{F}_{m,n,q}$ of the 3-modes interferometer fed with a $\vert 1,1,1 \rangle$ input state. (d) Diagram of the $N$-fold visibilities $V^{F}_{m,n,q}$ of the output fringe patterns $P^{F}_{m,n,q}$ for a $N=3$ interferometer with a $\vert 1,1,1 \rangle$ input state, compared with the classical bound $\Gamma_{m,n,q}^{cl}$. (e) Corresponding diagram for the 4-modes case.}
\label{fig:fig2}
\end{figure*}

\section{Results}

\subsection{Multi-arm interferometric schemes}

A multi-arm interferometer can be realized by cascading two multi-port beam splitters. In particular, a three-arm interferometer is build by the combination of two tritters and a four arm interferometer by the combination of two quarters. In our approach the integrated-optics multi-port device is devised as a 3D multi-waveguide directional coupler [Fig. \ref{fig:fig1} (a)-(b)], a structure in which the waveguides are brought close together for a certain interaction length and couple by evanescent field. The feasibility of such structures, in the case of three-ports (tritter), has already been demonstrated by femtosecond laser writing \cite{Kowa05, Suzu06}, albeit characterization has only been performed in a classical framework. Note that in these compact multi-waveguide structures the interaction between the different arms happens simultaneously, without the decomposition into cascaded two-mode beam splitters. This is made possible by the 3D capabilities of femtosecond laser micromachining, thus relaxing the strict requirements on path-length control of alternative approaches.

The symmetric configuration of a tritter can be easily obtained by adopting a triangular geometry, as shown in Fig. \ref{fig:fig1} (c). In this configuration it is possible to obtain equal coupling coefficients, so that a single photon entering in one input port has the same probability of exiting from one of the three output ports. The symmetric configuration of a quarter can be achieved by adopting two possible solutions. In the first one, the four modes are directly coupled, and the symmetric condition can be obtained by appropriately tuning the interaction length [Fig. \ref{fig:fig1} (d)]. Alternatively, an indirect geometry may be exploited by coupling waveguides 1-4 through an ancillary mode, as shown in Fig. \ref{fig:fig1} (e), and without any direct coupling between them.

\subsection{Output states} - We start by considering the action of a single tritter and of a single quarter. The action of these devices on an input state $\vert \psi \rangle$ is expressed by a unitary matrix $\mathcal{U}^{(III)}$ $[\mathcal{U}^{(IV)}]$, which maps the input field operators $a^{\dag}_{i}$ to the output field operators $b^{\dag}_{i}$ according to $b^{\dag}_{i} = \sum_{i,j} \mathcal{U}^{(k)}_{ij} a^{\dag}_{j}$, with $k=III,IV$ \cite{Zuko97,Camp00,Tich10} (see Supplementary Material). Let's consider the Fock state $\vert 1,1,1 \rangle$, where $\vert i,j,l \rangle = \vert i \rangle_{\mathbf{k}_{1}} \vert j \rangle_{\mathbf{k}_{2}} \vert l \rangle_{\mathbf{k}_{3}}$ is the input state in the tritter.  By applying $\mathcal{U}^{(III)}$ we obtain the output state:
\begin{equation}
\vert 1,1,1 \rangle \rightarrow c_{\{1,1,1\}} \vert 1,1,1 \rangle + c_{\{3,0,0\}} \vert \{3,0,0\} \rangle.
\end{equation}
Here $c_{1,1,1}=-e^{\imath 2 \pi/3}/\sqrt{3}$, $c_{\{3,0,0\}}=e^{\imath 4 \pi/3} \sqrt{2/3}$, and $\vert \{i,j,l\} \rangle$ is the symmetric superposition of three-photon states where $(i,j,l)$ photons exit in the three output ports. A similar result is obtained for a quarter fed with a $\vert 1,1,1,1 \rangle=\vert 1 \rangle_{\mathbf{k}_{1}} \vert 1 \rangle_{\mathbf{k}_{2}} \vert 1 \rangle_{\mathbf{k}_{3}} \vert 1 \rangle_{\mathbf{k}_{4}}$ state:
\begin{equation}
\begin{aligned}
\vert 1,1,1,1 \rangle &\rightarrow c_{\{1,1,1,1\}} \vert 1,1,1,1 \rangle + c_{\{2,2,0,0\}} \vert \{2,2,0,0\} \rangle + \\ &+ c_{\{4,0,0,0\}} \vert \{4,0,0,0 \} \rangle,
\end{aligned}
\end{equation}
where $c_{1,1,1,1}=1/2$, $c_{\{2,2,0,0\}}=\sqrt{6}/4$, $c_{\{4,0,0,0\}}=-\sqrt{6}/4$.
In both cases, we observe that some terms of the output states are suppressed due to quantum interference. They correspond in particular to the contributions $\{2,1,0\}$ in the tritter case, $\{3,1,0,0\}$ and $\{2,1,0,0\}$ in the quarter one. This feature is a $N$-mode analogue of the Hong-Ou-Mandel bosonic coalescence effect \cite{Hong87,Ou99,Tich10}. The two devices can be exploited to generate maximally-entangled N00N states in a post-selected configuration \cite{Pryd03}.

\subsection{N-modes interferometry} - The 3D, $N$-port structure of tritter and of quarter devices can be exploited to implement an integrated $N$-modes interferometer. This can be realized by a chain of two subsequent multiport beam-splitter, leading to a generalized Mach-Zehnder structure [see Fig. \ref{fig:fig1} (f)-(g)]. The phase shifting could be introduced for instance by adopting a microfluidic channel as reported in Ref. \cite{Cres10,Cres12}. Let us now consider the 3-modes system obtained with two tritter devices, and the action of a relative phase shift $\phi$ in the optical mode $\mathbf{k}_{3}$ inside the interferometer, which is described by the operator $\mathcal{U}_{\phi}= \exp{(-\imath n_{3} \phi)}$, being $n_{3}$ the photon number operator for mode $\mathbf{k}_{3}$. The output probability distributions $P_{m,n,q}(\phi)$ corresponding to the detection of $(m,n,q)$ photons in the three output ports, are obtained by the overall evolution of the interferometer $\mathcal{U}^{(III)} \mathcal{U}_{\phi} \mathcal{U}^{(III)}$ acting on the input state. Let us consider again an input Fock state $\vert 1,1,1 \rangle$; the obtained fringe patterns $P^{\mathrm{F}}_{m,n,q}(\phi)$, symmetric for an index exchange $(m,n,q)$, are reported in Figs. \ref{fig:fig2} (a)-(c). We observe the presence of interferometric patterns presenting the sum of different harmonics up to $\cos(3 \phi)$. Furthermore we note that the $P^{\mathrm{F}}_{2,1,0}(\phi)$ term presents a sub-Rayleigh $\lambda/3$ behavior with a unitary visibility. These results suggest that the output state of the interferometer, for a $\vert 1,1,1 \rangle$ input state, presents nonclassical features. Similar results can be obtained when a 4-modes interferometer is fed with a $\vert 1,1,1,1 \rangle$ input state [Fig. \ref{fig:fig1} (g)] (see Supplementary Material).

%
%
\begin{figure*}[ht!]
\includegraphics[width=0.97\textwidth]{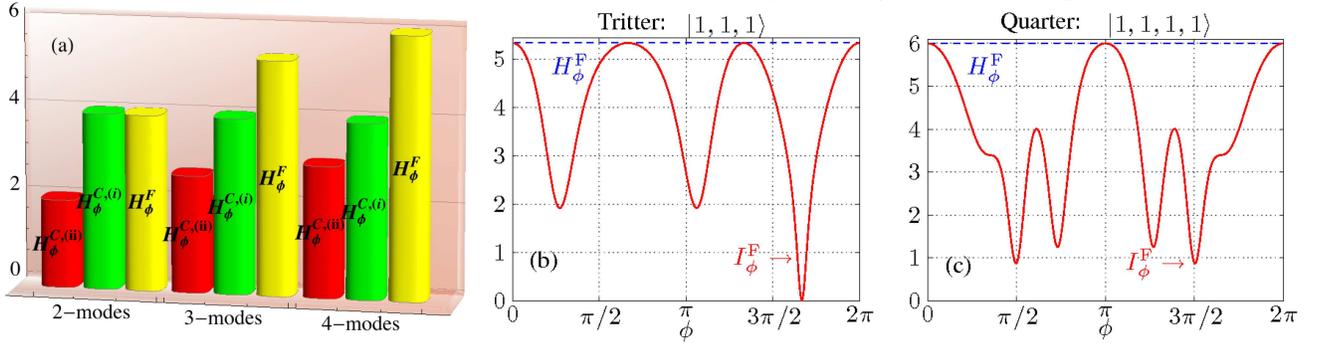}
\caption{(a) Comparison of the QFIs for the $N=2$, $N=3$, $N=4$ interferometers fed with a Fock state $H_{\phi}^{\mathrm{F}}$, a coherent state with an external phase reference beam $H_{\phi}^{\mathrm{C,(i)}}$, and a coherent state without an external phase reference beam $H_{\phi}^{\mathrm{C,(ii)}}$. (b)-(c) QFI $H_{\phi}^{\mathrm{F}}$ and FI $I_{\phi}^{\mathrm{F}}$ for (b) a 3-modes interferometer with a $\vert 1,1,1 \rangle$ input state, and (c) for a 4-modes interferometer with an input $\vert 1,1,1,1 \rangle$ state. The optimal working points are obtained when $I_{\phi}^{\mathrm{F}}=H_{\phi}^{\mathrm{F}}$.}
\label{fig:fig3}
\end{figure*}

\subsection{Nonclassicality criterion for sub-Rayleigh fringe patterns} - In order to formally address the nonclassicality of this setup we have extended the criterion proposed by Afek \textit{et al.} in Ref. \cite{Afek10a} from a 2-modes to a 3-modes interferometer. This criterion sets an upper bound $\Gamma^\mathrm{cl}_{m,n,q}$ for classical states on the $N$-fold visibilites of the Fourier expansion of the fringe pattern distributions $P_{m,n,q}(\phi) = \sum_{k=0}^{\infty} A_{k} \cos(k \phi - \delta_{k})$, where $N=m+n+q$. The visibilities are defined as the ratio between the Fourier coefficient $A_{m,n,q}$ of the term oscillating as $N\phi$ and the constant coefficient $A_0$, i.e. $V_{m,n,q}=\vert {A_{m,n,q}}/{A_0} \vert$. The classical bound for $V_{m,n,q}$ is obtained by calculating the probability distribution of having $(m,n,q)$ photons in the output modes $(1,2,3)$ respectively, by feeding the interferometer with a classical coherent state: $\vert \alpha_1, \alpha_2, \alpha_3 \rangle$, where $\alpha_i=|\alpha_i|e^{\imath \theta_i}$. Then, the $N$-fold visibility $V_{m,n,q}$ is maximized with respect to $\vert \alpha_i\vert$ and $\theta_i$, leading to $\max V_{m,n,q} = \Gamma_{m,n,q}^{\mathrm{cl}}$. In general, it can be shown that for any classical state $V_{m,n,q}\leq \Gamma^\mathrm{cl}_{m,n,q}$, since any classical state can be expanded in the coherent states basis according to $\rho_C=\int d^2\alpha \, P(\alpha) \vert \alpha \rangle \langle \alpha \vert$ with a well behaved $P(\alpha)$. The corresponding $N$-fold visibilities $V^{\mathrm{F}}_{m,n,q}$ for the input Fock state $\vert 1,1,1 \rangle$ are obtained from the probability distributions $P^{\mathrm{F}}_{m,n,q}$ shown in Figs. \ref{fig:fig2} (a)-(c), leading to higher visibilities than the classical limits, except for the $P^{F}_{1,1,1}$ case [Fig. \ref{fig:fig2} (d)]. The same criterion can be extended to a $N$-modes interferometer. We then repeated the same analysis for the $N=4$ case when the interferometer is built by two subsequent quarter devices [Fig. \ref{fig:fig1} (g)], showing the presence of nonclassical behaviors in the fringe patterns [see Fig. \ref{fig:fig2} (e)].

\subsection{Phase estimation} - The present interferometric configuration can be adopted to perform a phase estimation protocol. In this context, the aim is to measure an unknown phase shift $\phi$ introduced in an interferometer with the best possible precision by probing the system with a $N$-photon state, and by measuring the resulting output state. The classical limit is provided by the SQL, which sets a lower bound to the minimum uncertainty $\delta \phi_{\mathrm{SQL}} \geqslant 1/\sqrt{M N}$ which can be obtained on $\phi$ by exploiting classical $N$-photon states on two modes and $M$ repeated measurements \cite{Ou97}. Recently, it has been shown that the adoption of quantum states can lead to a better scaling with $N$, setting the ultimate precision to $\delta \phi_{\mathrm{HL}} \geqslant 1/(\sqrt{M} \, N)$, corresponding to the Heisenberg limit \cite{Giov04b,Giov06,Giov11}. Hereafter, we show that the present integrated technology can lead to a sub-SQL performance in the estimation of an optical phase, exploiting multi-mode interferometry.

\subsection{Quantum Fisher information} - In order to characterize the present $3$-modes interferometer fed with a $\vert 1,1,1 \rangle$ input state, and analogously the $4$-modes case, we need to determine the quantum Fisher information (QFI) $H^{\mathrm{F}}_{\phi}$ of the output state \cite{Hels76,Brau96}. This quantity sets the maximum amount of information which can be extracted on the phase $\phi$ from a state $\varrho_{\phi}$ according to the quantum Cram{\'e}r-Rao (QCR) bound: $\delta \phi \geq (M H_{\phi})^{-1/2}$ \cite{Hels76}. The classical limit is provided by the quantum Fisher information $H_{\phi}^{\mathrm{C}}$ when a coherent state $\vert \alpha_{1},\alpha_{2},\alpha_{3} \rangle$ is injected into the interferometer. Note that the comparison between the performances achievable with an input coherent state and an input $\vert 1,1,1 \rangle$ Fock state must be performed for the same number of photons impinging onto the phase shifter. Furthermore, for a coherent state input two different cases can be identified. {\bf (i)} If an external reference beam, providing an absolute reference frame for the optical phase, is available at the measurement stage, the QFI is calculated for a pure $\vert \alpha_{1}, \alpha_{2},\alpha_{3} \rangle$ input state: $H_{\phi}^{\mathrm{C},(i)}$. {\bf (ii)} If no reference frame is available at the measurement stage (such as for photon-counting detection), one needs to average the input state on a random phase shift $\theta$ common to all input modes leading to $H_{\phi}^{\mathrm{C},(ii)}$. In absence of an external reference beam, an input state $\varrho$ has to be replaced with the phase-averaged state $\varrho' = (2 \pi)^{-1}\int_0^{2\pi} d\theta \, \mathcal{U}_{\theta}^{1} \mathcal{U}_{\theta}^{2} \mathcal{U}_{\theta}^{3} \varrho \mathcal{U}_{\theta}^{1\, \dag} \mathcal{U}_{\theta}^{2 \, \dag} \mathcal{U}_{\theta}^{3 \, \dag}$, where $\mathcal{U}_{\theta}^{i}$ is the phase shift operator for mode $\mathbf{k}_{i}$. The two conditions (i) and (ii) are equivalent for the $\vert 1,1,1 \rangle$ probe, since this state has a fixed number of photons \cite{Jarz11}. We then evaluated the three quantities $H_{\phi}^{\mathrm{C},(i)},H_{\phi}^{\mathrm{C},(ii)},H_{\phi}^{\mathrm{F}}$, and obtained that the adoption of a $\vert 1,1,1 \rangle$ probe state leads to quantum improved performances. The same result is found for the $4$-modes case, as shown in Fig. \ref{fig:fig3} (a). We note that, while $H^{\mathrm{C},(i)}_{\phi}$ is fixed, the QFI achievable with a Fock state input increases with the number of modes, leading progressively to a greater advantage in phase estimation protocols. Furthermore, no post-selection is needed to generate the required probe state \cite{Resc07}.
	
\subsection{Achieving the optimal bound} - We now show that the QCR bound provided by $H_{\phi}^{\mathrm{F}}$ can be achieved by adopting a feasible and practical choice of the measurement setup, consisting in a photon-counting apparatus recording the number of output photons on each mode. The detection apparatus can be implemented by splitting each output mode in three parts by means of a chain of beam-splitters, and by placing a single-photon detector on each part. The occurrence of 1, 2 or 3 simultaneous clicks of the detectors on the same mode corresponds to the detection of 1, 2, 3 photons. For a fixed choice of the measurement setup, the amount of information which can be extracted on $\phi$ is provided by the Cram{\'e}r-Rao (CR) bound $\delta \phi \geq (M I_{\phi})^{-1/2}$ \cite{Hels76}, where $I_{\phi}$ is the Fisher information of the output probability distribution of the measurement outcomes. The results for $I^{\mathrm{F}}_{\phi}$ with the $\vert 1,1,1 \rangle$ input state and photon-counting measurements are reported in Fig. \ref{fig:fig3} (b). By comparing the trend of $I^{\mathrm{F}}_{\phi}$ with the corresponding QFI $H_{\phi}^{\mathrm{F}}$, we observe that the ultimate precision given by the QCR bound can be achieved with this choice of the measurement apparatus for $\phi=0,2 \pi/3,4 \pi/3$. An analogous result is found for the $N=4$ case, where the optimal working points are now $\phi=0,\pi$ [Fig. \ref{fig:fig3} (c)].

\subsection{Adaptive protocol} 
\begin{figure}[ht!]
\centering
\includegraphics[width=0.45\textwidth]{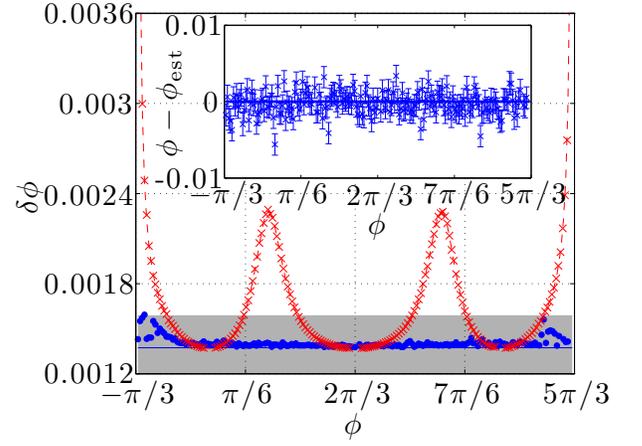}
\caption{Results for the phase estimation error $\delta \phi$ of a numerical simulation with $M=10^{5}$ repeated measurements with the three modes interferometer of Fig. \ref{fig:fig1} (f). Blue circular points: $\delta \phi$ for the adaptive protocol as a function of $\phi$. Blue solid line: QCR bound given by $H_{\phi}^{\mathrm{F}}$. Red cross points: $\delta \phi$ for the non-adaptive protocol. Red dashed line: Fisher information $I_{\phi}^{\mathrm{F}}$, providing the bound for the non-adaptive strategy. Shaded region corresponds to sub-SQL performances achievable with quantum resources. Inset: plot of the difference between the true value $\phi$ and the estimated value $\phi_{\mathrm{est}}$ obtained with the adaptive protocol.}
\label{fig:fig4}
\end{figure}

- The obtained $\phi$-dependence of the Fisher information $I_{\phi}$ suggests that an adaptive protocol \cite{Naga88} is necessary to obtain optimal performances in the full phase range, that is, to saturate the QCR bound for all values of $\phi$. To this purpose, we consider the adoption of a three-step adaptive strategy where the first two steps of the protocol are performed to obtain a rough estimate $\phi_{\mathrm{r}}$ of the phase $\phi$ (more details on the protocol can be found in the Supplementary information). The amount of measurements performed in these steps is a small fraction of the overall resources $M$, namely $M_{1}=M_{2}=\sqrt{M}$. In the first step, a rough estimate of the phase is obtained up to a two-fold degeneracy due to the symmetry of the interferometric fringes ($\phi$,$4 \pi/3-\phi$). The second step is exploited to remove this degeneracy for the estimate $\phi_{\mathrm{est}}$. Finally, the third step consisting of $M_{3}=M-M_{1}-M_{2}$ measurements is performed by sending the $\vert 1,1,1 \rangle$ input state, and the system is tuned to operate in the optimal regime ($\phi+\psi \simeq 2 \pi/3$) by means of an additional phase shift $\psi$. At each step of the protocol, the measurement outcomes are analyzed by a Bayesian approach and assuming no a-priori knowledge on $\phi$ \cite{Pezz07}. The results of the numerical simulation for $M=10^{5}$ are reported in Fig. \ref{fig:fig4}, together with the results for a numerical simulation of a one step non-adaptive strategy. 
%
%
We observe that in the latter case, the uncertainty $\delta \phi$ associated to the estimation process resembles closely the CR bound provided by the Fisher information $I_{\phi}^{\mathrm{F}}$. A better result is obtained with the adaptive strategy, showing that the quantum Fisher information is achieved, leading to sub-SQL performances in the full phase interval. The choice of the Bayesian approach leads to an unbiased estimation process, i.e., the estimated phase $\phi_{\mathrm{est}}$ converges to the true value $\phi$ (see inset of Fig. \ref{fig:fig4}), and the error on the estimation process can be directly retrieved from the output distribution for $\phi$ \cite{Kris11}.

\subsection{Two parameter phase estimation} 
\begin{figure*}[ht!]
\centering
\includegraphics[width=0.85\textwidth]{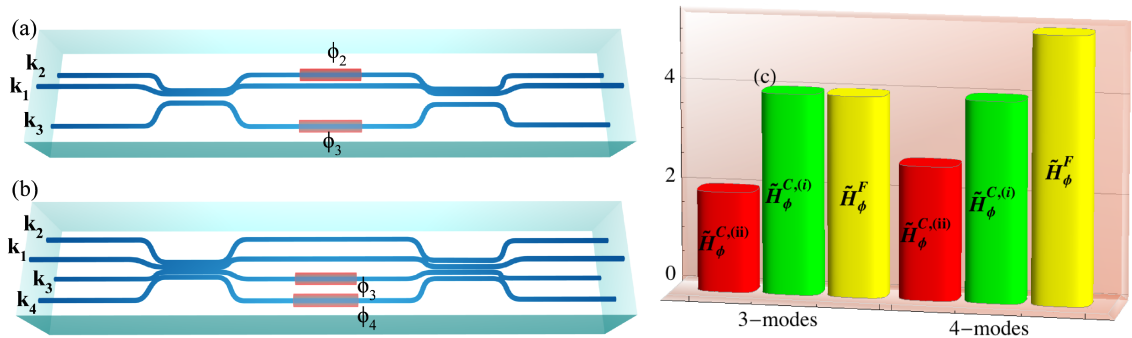}
\caption{(a)-(b) Schemes for two parameters phase estimation with the $3$- and $4$-modes integrated interferometers. (c) Comparison between the effective quantum Fisher informations ($\tilde{H}_{\phi_{\mu}}^{C,(i)}$, $\tilde{H}_{\phi_{\mu}}^{C,(ii)}$, $\tilde{H}_{\phi_{\mu}}^{F}$) for the two-parameters problem when the interferometer is fed with Fock states or coherent states.}
\label{fig:two_parameters}
\end{figure*}

- Let us now consider a different scenario. The $3$-modes interferometer built with two cascaded tritters proposed in this paper may be adopted to perform a two parameters estimation process, consisting of the simultaneous measurement of two optical phases. In this case, the reference $\phi_{\mathrm{ref}}$ is provided by the optical mode $\mathbf{k}_{1}$, and the phases to be measured, $\phi_{2}$ and $\phi_{3}$, correspond respectively to the optical modes $\mathbf{k}_{2}$ and $\mathbf{k}_{3}$ [see Fig. \ref{fig:two_parameters} (a)]. In order to evaluate the maximum precision achievable in the two parameter problem, it is necessary to extend the concept of quantum Fisher information to the multiparameter case \cite{Pari09}. Indeed, it is possible to define a quantum Fisher information matrix (QFIM) $\mathbf{H}_{\mu \nu}$ corresponding to the set of parameter ${\bm \lambda} = (\lambda_{1},\ldots,\lambda_{n})$, which in the pure state case $\vert \psi_{{ \bm \lambda}} \rangle = e^{-\imath \sum_{\mu} G_{\mu} \lambda_{\mu}} \vert \psi_{0} \rangle$ is defined in terms of the set of generators $\mathbf{G} = (G_{1},\ldots,G_{n})$ for the parameters ${\bm \lambda}$. The error on the single parameter $\lambda_{\mu}$ for a fixed value of the other parameters $\lambda_{\nu}$, with $\nu \neq \mu$ is bounded by the inequality: $\delta \lambda_{\mu} \geq [(\mathbf{H}^{-1})_{\mu \mu}/M]^{1/2}.$ When performing the simultaneous estimation of the set of parameters ${\bm \lambda}$, the sum of the variances is bounded by the multiparameter Cramer-Rao inequality: $\sum_{\mu} \mathrm{Var}(\lambda_{\mu}) \geq \mathrm{Tr}[\mathbf{H}^{-1}]/M$.

%

We now consider as input state a coherent state $\vert \alpha \rangle$ with $\langle n \rangle = 3$, which defines the standard quantum limit for a $N=3$ photon probe. As for the single parameter scenario, it is necessary to evaluate the quantum Fisher information matrices both (i) in presence of an external phase reference $\mathbf{H}^{C,(i)}_{\phi_{2},\phi_{3}}$ and (ii) in absence of an external phase reference $\mathbf{H}^{C,(ii)}_{\phi_{2},\phi_{3}}$. The corresponding bounds for the sensitivities after $M$ measurements in the two cases are given by $\delta \phi^{C,(i)}_{\mu} \geq (M \tilde{H}_{\phi_{\mu}}^{C,(i)})^{-1/2}$ and $\delta \phi^{C,(ii)}_{\mu} \geq (M \tilde{H}_{\phi_{\mu}}^{C,(ii)})^{-1/2}$. When a three photon state $\vert 1,1,1 \rangle$ is injected into the interferometer, the bounds for the parameters $\phi_{2}$ and $\phi_{3}$ are the same for both cases (i) and (ii) leading to $\delta \phi^{F}_{\mu} \geq (M \tilde{H}_{\phi_{\mu}}^{F})^{-1/2}$. The same analysis may be performed in the case of a $4$-modes interferometer, fed with a coherent state of $\langle n \rangle = 4$ photons or with a Fock state $\vert 1,1,1,1 \rangle$, where the two parameters are now the phases $\mathbf{\phi}_{3}$ and $\mathbf{\phi}_{4}$ on modes $\mathbf{k}_{3}$ and $\mathbf{k}_{4}$ [see Fig. \ref{fig:two_parameters} (b)]. The results are shown in Fig. \ref{fig:two_parameters} (c), showing that in absence of a phase reference the adoption of Fock probe states can lead to quantum enhanced performances in the measurement of two optical phases. Furthermore, in full analogy with the one parameter case, a greater advantange with respect to the classical strategies may be progressively achieved by increasing the number of modes.

In the single parameter case, the quantum Cramer-Rao bound can always be asymptotically achieved \cite{Naga88} performing a suitable measurement and choosing the right estimator. On the contrary, in the multiparameter case the bound for the statistical errors defined by the quantum Fisher information matrix is not in general achievable \cite{Crow12,Geno12}. This depends on the fact that the optimal measurements for the individual parameters may not be compatible observables. A necessary condition for the achievability of the multiparameter quantum Cramer-Rao bound is then given by the weak commutativity condition: $\mathrm{Tr}[\rho_{\bm \lambda}[L_{\mu},L_{\nu}]]=0$. Here, the $L_{\mu}$ operators are the symmetric logarithmic derivatives (SLD), which define the optimal measurement and estimators for the individual parameters \cite{Gill11}. In our case, it can be shown that (see Supplementary material) for the $3$-mode interferometer injected by the $\vert 1,1,1 \rangle$ input state the operators $\{L_{2},L_{3}\}$ for $\{\phi_{2},\phi_{3}\}$ commute, thus satisfying the necessary condition to achieve the quantum Cramer-Rao bound. The same result holds for the $4$-mode case. The next step to be investigated is the identification of suitable measurements and estimators.

\section{Discussions}

The analysis performed in this work represents the first step in the investigation of integrated quantum technology in view of the realization of multiport optical beam splitters enabling novel multiphoton sensing schemes based on 3D interferometers. We investigated the adoption of this technology to perform phase estimation protocols leading to quantum-enhanced performances. We provided and simulated a full protocol for sub-SQL phase measurements, by exploiting Fock input states and photon-counting detection, thus not requiring any post-selection for the generation of the probe state. We also discussed the application of the same multimode structure for multiparameter estimation purposes. The present technology is expected to lead to the development of new phase estimation protocols able to reach Heisenberg-limited performances \cite{Pryd03} and to open a new scenario for the simultaneous measurement of more than one optical phase. Indeed, the present approach can be adopted as an accessible test bench to investigate theoretically and experimentally the still unexplored scenario of multiparameter estimation. Further perspectives may lead to the application of this multiport splitters in other contexts, such as quantum simulations \cite{Lloy96}, linear-optical computing \cite{Aaro10} and nonlocality tests \cite{Gruc11}.

\section{Acknowledgements} 
This work was supported by FIRB-Futuro in Ricerca HYTEQ, ERC-Starting Grant 3D-QUEST.

%

\end{document}


\title{Supplementary material: Quantum interferometry with three-dimensional geometry}

\author{Nicol\`o Spagnolo}
\affiliation{Dipartimento di Fisica, Sapienza Universit\`{a} di Roma, Piazzale Aldo Moro 5, I-00185 Roma, Italy}
\author{Lorenzo Aparo}
\affiliation{Dipartimento di Fisica, Sapienza Universit\`{a} di Roma, Piazzale Aldo Moro 5, I-00185 Roma, Italy}
\author{Chiara Vitelli}
\affiliation{Center of Life NanoScience @ La Sapienza, Istituto Italiano di Tecnologia, Viale Regina Elena, 255, I-00185 Roma, Italy}
\affiliation{Dipartimento di Fisica, Sapienza Universit\`{a} di Roma, Piazzale Aldo Moro 5, I-00185 Roma, Italy}
\author{Andrea Crespi}
\affiliation{Istituto di Fotonica e Nanotecnologie, Consiglio Nazionale delle Ricerche (IFN-CNR), Piazza Leonardo da Vinci, 32, I-20133 Milano, Italy}
\affiliation{Dipartimento di Fisica, Politecnico di Milano, Piazza Leonardo da Vinci, 32, I-20133 Milano, Italy}
\author{Roberta Ramponi}
\affiliation{Istituto di Fotonica e Nanotecnologie, Consiglio Nazionale delle Ricerche (IFN-CNR), Piazza Leonardo da Vinci, 32, I-20133 Milano, Italy}
\affiliation{Dipartimento di Fisica, Politecnico di Milano, Piazza Leonardo da Vinci, 32, I-20133 Milano, Italy}
\author{Roberto Osellame}
\affiliation{Istituto di Fotonica e Nanotecnologie, Consiglio Nazionale delle Ricerche (IFN-CNR), Piazza Leonardo da Vinci, 32, I-20133 Milano, Italy}
\affiliation{Dipartimento di Fisica, Politecnico di Milano, Piazza Leonardo da Vinci, 32, I-20133 Milano, Italy}
\author{Paolo Mataloni}
\affiliation{Dipartimento di Fisica, Sapienza Universit\`{a} di Roma, Piazzale Aldo Moro 5, I-00185 Roma, Italy}
\affiliation{Istituto Nazionale di Ottica (INO-CNR), Largo E. Fermi 6, I-50125 Firenze, Italy}
\author{Fabio Sciarrino}
\affiliation{Dipartimento di Fisica, Sapienza Universit\`{a} di Roma, Piazzale Aldo Moro 5, I-00185 Roma, Italy}
\affiliation{Istituto Nazionale di Ottica (INO-CNR), Largo E. Fermi 6, I-50125 Firenze, Italy}

\begin{abstract}
In this Supplementary material we provide more details on the results presented in the paper. We first report the unitary evolution matrices which describe the action of tritter and quarter devices. Then, we analyze the multiarm interferometers obtained by cascading two tritter (quarter) devices and deriving the probability distributions associated to the output photon number statistics for the input state $\vert 1,1,1 \rangle$ ($\vert 1,1,1,1 \rangle$). We discuss more in detail the analyzed phase estimation protocol based on the 3-dimensional interferometric architecture obtained with multiport devices. Finally, we perform a first analysis on the two parameter estimation problem.
\end{abstract}

\maketitle

\section{Tritter and quarter unitary evolution}

The basic element of a tritter (quarter) is a directional coupler (DC) in which 3 (4) waveguides are brought close together and coupled by evanescent field. A DC is described by transmission coefficients $T_{mn}=|T_{mn}|e^{\imath \phi_{mn}}$, giving the probability amplitude of the photon exiting in the same optical mode ($m = n$), and the coupling between different waveguides ($m \neq n$). In the symmetric case we have $\vert T_{mn} \vert = 1/\sqrt{N}$ for any $(m,n)$, where $N$ is the number of waveguides in the DC. The time evolution of the field operators is obtained as $b^{\dag}_{i} = \sum_{i,j} \mathcal{U}^{(k)}_{ij} a^{\dag}_{j}$, where $k=III,IV$, $\{ b^{\dag}_{i}\}$ are the output modes operators, and $\{a^{\dag}_{i}\}$ are the input modes operators. The matrix $\mathcal{U}^{(III)}$ in the symmetric case is obtained by imposing the unitariety condition on the transformation induced by the linear coupling coefficients $T$ and $R$, leading to:
\begin{equation}
\mathcal{U}^{(III)} =\frac{1}{\sqrt 3}\begin{pmatrix} 1 & e^{\frac{\imath 2 \pi}{3}} & e^{\frac{\imath2\pi}{3}}\\ e^{\frac{\imath 2\pi}{3}} & 1 & e^{\frac{\imath2\pi}{3}}\\ e^{\frac{\imath 2\pi}{3}} & e^{\frac{\imath 2\pi}{3}} & 1 \end{pmatrix}.
\label{matrix5050}
\end{equation}
Following the same procedure the unitary matrix which describes the evolution in the symmetric quarter device is given by:
\begin{equation}
\mathcal{U}^{(IV)} =\frac{1}{2}\begin{pmatrix} 1 & -1 & -1 & -1\\ -1 & 1 & -1 & -1\\ -1 & -1 & 1 &-1\\ -1 & -1 & -1 &1 \end{pmatrix}.
\end{equation}
As shown in Ref. \cite{Zuko97}, there is no unique choice for the matrices $\mathcal{U}^{(III)}$ and $\mathcal{U}^{(IV)}$. In the tritter case, there is only one equivalence class for the matrices $\mathcal{U}^{(III)}$, that is, all the allowed transformations are equivalent up to phase shifts inserted in the input or in the output modes. In the quarter case there is a set of equivalence classes for the matrices $\mathcal{U}^{(IV)}$, parametrized by a continuous parameter $\theta \in [0, 2 \pi)$.

\section{Interference fringes}

A 3-photon Mach-Zehnder interferometer is obtained by cascading two balanced tritters, and considering the action of a phase shift $\phi$ on mode $\mathbf{k}_{3}$ described by the unitary evolution $\mathcal{U}_{\phi}=\exp(-\imath n_{3} \phi)$. The output state $\vert \psi_{\mathrm{out}} \rangle$ of the interferometer corresponding to an input state $\vert \psi_{\mathrm{in}} \rangle$ is obtained as $\vert \psi_{\mathrm{out}} \rangle = \mathcal{U} \vert \psi_{\mathrm{in}} \rangle$, where $\mathcal{U} = \mathcal{U}^{(III)} \mathcal{U}_{\phi} \mathcal{U}^{(III)}$ represents the overall evolution of the system. By feeding the interferometer with the input state $\vert 1,1,1\rangle$ the output probability distributions of obtaining $(m,n,q)$ photons on the three output modes take the form:
\begin{equation}
P_{1,1,1}^{\mathrm{F}}(\phi)=\frac{29}{81} - \frac{24}{81} \cos (\phi+\pi/3) - \frac{12}{81} \cos ( 2 \phi - \pi/3) + \frac{16}{81} \cos (3 \phi),
\end{equation}
\begin{equation}
\begin{aligned}
P_{2,1,0}^{\mathrm{F}}(\phi)=\frac{4}{81} \left[1 - \cos \left( 3 \phi \right) \right],
\end{aligned}
\end{equation}
\begin{equation}
P^{\mathrm{F}}_{3,0,0}(\phi)=\frac{28}{243} - \frac{24}{243}  \cos ( \phi - 2 \pi/3) + \frac{12}{243} \cos ( 2\phi - \pi/3) +\frac{8}{243} \cos (3 \phi).
\end{equation}
All the probability distributions $P_{m,n,q}^{\mathrm{F}}$ are symmetric with respect to an exchange of the output indices $(m,n,q)$.

By an analogous procedure, the output probability distributions for the 4-modes interferometer, obtained by two cascaded quarter devices and fed with the input state $\vert 1,1,1,1 \rangle$, read:
\begin{equation}
P_{1,1,1,1}^{\mathrm{F}}(\phi)=\frac{167}{512}+\frac{168}{512} \cos \phi + \frac{108}{512} \cos ( 2 \phi) + \frac{24}{512} \cos (3 \phi )+\frac{45}{512} \cos ( 4 \phi ),
\end{equation}
\begin{equation}
\begin{aligned}
P_{2,2,0,0}^{\mathrm{F}}(\phi)=\frac{47+60\cos \phi + 21 \cos \left( 2 \phi\right)\sin^4 (\phi/2)}{128},
\end{aligned}
\end{equation}
\begin{equation}
\begin{aligned}
P_{3,1,0,0}^{\mathrm{F}}(\phi)=\frac{3\sin^4 \left( \phi\right)}{128},
\end{aligned}
\end{equation}
\begin{equation}
\begin{aligned}
P_{4,0,0,0}^{\mathrm{F}}(\phi)=\frac{3\left( 5+ 3\cos \phi \right)\sin^6(\phi/2)}{64},
\end{aligned}
\end{equation}
\begin{equation}
\begin{aligned}
P_{2,0,1,1}^{\mathrm{F}}(\phi)=\frac{[ 17+15 \cos (  2 \phi )]\sin^2 \phi}{256}.
\end{aligned}
\end{equation}
As for the 3-mode interferometer fed with the $\vert 1,1,1 \rangle$ state, the output probability distributions $P_{m,n,q,p}^{\mathrm{F}}$ of the 4-mode case are symmetric with respect to an exchange of the output indices $(m,n,q,p)$.

\section{Phase estimation with 3-mode and 4-mode interferometers}

Let us consider the 3-mode and the 4-mode interferometers obtained by cascading two tritter (quarter) devices, fed with the input state $\vert 1,1,1 \rangle$ ($\vert 1,1,1,1 \rangle$). We calculate the quantum Fisher information associated to the adopted probe state, and the classical Fisher information obtained with photon-counting detection. We then discuss an adaptive 3-step phase estimation protocol able to achieve optimal performance for an unknown phase shift lying in the full phase interval, that is, able to saturate the quantum Fisher information of the scheme.

\subsection{Quantum Fisher information}

In the context of parameter estimation, the quantum Fisher information $H_{\phi}$ represents the maximum amount of information which can be extracted on the phase $\phi$ from a family of states $\varrho_{\phi}$, according to the quantum Cram{\'e}r-Rao bound: $\delta \phi \geq (M H_{\phi})^{-1/2}$. When the family of states $\varrho_{\phi}$ is obtained from unitary evolution $\mathcal{U}_{\phi} = \exp(-\imath G \phi)$ of an input pure state $\vert \psi_{0} \rangle$ as $\varrho_{\phi} = \mathcal{U}_{\phi} \vert \psi_{0} \rangle \langle \psi_{0} \vert \mathcal{U}_{\phi}^{\dag}$, the quantum Fisher information is evaluated as $H_{\phi} = 4 \langle \psi_{0} \vert (\Delta G)^{2} \vert \psi_{0} \rangle$ \cite{Pari09}. In the case of the 3-mode interferometer fed with a $\vert 1,1,1 \rangle$ input state, the quantum Fisher information reads:
\begin{equation}
H^{\mathrm{F}}_{\phi} = 4 \langle 1,1,1 \vert \mathcal{U}^{(III) \, \dag} (\Delta n_{3})^{2} \mathcal{U}^{(III)} \vert 1,1,1 \rangle = \frac{16}{3},
\end{equation}
where $n_{3}=G$ is the generator of the phase shift on mode $\mathbf{k}_{3}$. In the case of the 4-mode interferometer fed with a $\vert 1,1,1,1 \rangle$ input state, the quantum Fisher information reads:
\begin{equation}
H^{\mathrm{F}}_{\phi} = 4 \langle 1,1,1,1 \vert \mathcal{U}^{(IV) \, \dag} (\Delta n_{4})^{2} \mathcal{U}^{(IV)} \vert 1,1,1,1 \rangle = 6,
\end{equation}
where $n_{4}=G$ is the generator of the phase shift on mode $\mathbf{k}_{4}$.

\subsection{Classical Fisher information}

When fixing the detection apparatus to the set of POVM operators $\{ \Pi_{x} \}$, the maximum amount of information which can be extracted on the phase $\phi$ from a family of states $\varrho_{\phi}$ measured by $\{ \Pi_{x} \}$ is quantified by the Cram{\'e}r-Rao bound $\delta \phi \geq (M I_{\phi})^{-1/2}$, where $I_{\phi}$ is the classical Fisher information:
\begin{equation}
I_{\phi} = \sum_{x} p(x \vert \phi) [\partial_{\phi} \log p(x \vert \phi)]^{2},
\end{equation}
where $p(x \vert \phi)$ is the conditional probability distribution of the measurement outcomes. In the case of the 3-mode interferometer fed with a $\vert 1,1,1 \rangle$ input state, when we choose photon-counting $\Pi_{m,n,q} = \vert m,n,q \rangle \langle m,n,q \vert$ as the measurement apparatus, we obtain:
\begin{widetext}
\begin{equation}
\begin{aligned}
I_{\phi}^{\mathrm{F}}&=\frac{16}{9}\left(3\cos^2 \left(\frac{3\phi}{2}\right)+\frac{2 \sin ^2\left(\frac{3 \phi }{2}\right) \left(\sqrt{3} \sin \left(\frac{\phi }{2}\right)+\cos \left(\frac{\phi }{2}\right)+2 \cos \left(\frac{3 \phi
   }{2}\right)\right)^2}{-6 \sqrt{3} \sin (\phi )+3 \cos (2 \phi )+4 \cos (3 \phi )+6 \left(\sqrt{3} \sin (\phi )+1\right) \cos (\phi )+14}\right. \\
    &+ \left. \frac{\left(\sin (\phi )+\sin (2 \phi )-4 \sin (3 \phi )+\sqrt{3} \cos (\phi )-\sqrt{3} \cos (2 \phi )\right)^2}{12 \sqrt{3} \sin (\phi )-6 \sqrt{3} \sin (2 \phi )-12 \cos (\phi )-6 \cos (2 \phi )+16 \cos (3 \phi )+29} \right).
   \end{aligned}
\end{equation}
An analogous result is obtained for the 4-mode interferometer.
\end{widetext}

\subsection{Three-step bayesian adaptive protocol}

We have also developed and simulated an adaptive phase estimation protocol tailored to reach asymptotically the quantum Cram{\'e}r-Rao bound of the input state for all values of the phase $\phi$. We adopt a Bayesian approach \cite{Pezz07}, which is based on inverting the probability distribution of the measurement outcomes according to the Bayes formula. Bayesian analysis present the relevant properties of being asymptotically optimal and unbiased, that is, the estimated value converges to the true value of the parameter with uncertainty reaching the quantum limit for large measurements $M$. Furthermore, the output distribution for the phase $\phi$ presents a fast convergence to a Gaussian shape, and the error on the estimation process can be directly retrieved from the output distribution for $\phi$ \cite{Kris11}. In the case of the 3-mode interferometer under analysis, the probability distribution of $M$ repeated photon-counting measurement outcomes $P(\{ N^{(i)}_{1}, N^{(i)}_{2}, N^{(i)}_{3} \}_{i=1}^{M} \vert \phi)$ is inverted to obtain a conditional distribution for the phase $\phi$:
\begin{equation}
\label{eq:pcond2}
P(\phi \vert \{ N^{(i)}_{1}, N^{(i)}_{2}, N^{(i)}_{3} \}_{i=1}^{M}) = \frac{P(\phi)}{\mathcal{N}} P(\{ N^{(i)}_{1}, N^{(i)}_{2}, N^{(i)}_{3} \}_{i=1}^{M} \vert \phi),
\end{equation}
where $\mathcal{N}$ is a normalization constant, evaluated as:
\begin{equation}
\mathcal{N} = \int_{0}^{2 \pi} d\phi \; P(\{ N^{(i)}_{1}, N^{(i)}_{2}, N^{(i)}_{3} \}_{i=1}^{M} \vert \phi) P(\phi).
\end{equation}
Here, $P(\phi)$ is the phase probability distribution which represents the a-priori knowledge on the value of the phase $\phi$. Finally, the estimated value of the phase $\phi_{\mathrm{est}}$ and the associated variance $\sigma^{2}_{\mathrm{est}}$ are obtained from the conditional distribution $P(\phi \vert \{ N^{(i)}_{1}, N^{(i)}_{2}, N^{(i)}_{3} \}_{i=1}^{M})$ as:
\begin{eqnarray}
\label{eq:bayesian_1}
\phi_{\mathrm{est}} &=& \int_{\Omega} d \phi \; \phi \, P(\phi \vert \{ N^{(i)}_{1}, N^{(i)}_{2}, N^{(i)}_{3} \}_{i=1}^{M}),\\
\label{eq:bayesian_2}
\sigma^{2}_{\mathrm{est}} &=& \int_{\Omega} d \phi \; (\phi - \phi_{\mathrm{est}})^{2} \, P(\phi \vert \{ N^{(i)}_{1}, N^{(i)}_{2}, N^{(i)}_{3} \}_{i=1}^{M}),
\end{eqnarray}
where $\Omega$ is the phase interval. More specifically, we considered the case in which there is maximum ignorance on the value of the phase $\phi$, corresponding a uniform distribution $P(\phi)$ in the interval $\Omega = [-\pi/3,5 \pi/3)$.

The strategy analyzed in the paper is an adaptive protocol, since it is based on driving an additional phase $\psi$ according to the results of each step of the experiment. The protocol is divided in three main steps:

\begin{itemize}
\item[(I)] In a first step, we perform $M_{1} = \sqrt{M}$ measurements by sending a $\vert 1,0,0 \rangle$ state in the interferometer starting from a uniform a priori distribution $P(\phi) = (2 \pi)^{-1}$, expressing no knowledge on the phase. In this case, we obtain a rough estimate of $\phi_{\mathrm{r}}^{1}$ in terms of a new distribution $P^{(1)}(\phi \vert \{ N_{1}^{(i)}, N_{2}^{(i)}, N_{3}^{(i)} \}_{i=1}^{M_{1}})$ by exploiting a Bayesian approach. The obtained value of $\phi_{\mathrm{r}}^{1}$ still presents a two-fold degeneracy ($\phi_{\mathrm{r}}^{1,(a)}$,$\phi_{\mathrm{r}}^{1,(b)}$) due to the periodicity of the fringe patterns.
\item[(II)] In a second step, we perform $M_{2} = \sqrt{M}$ measurements by sending a $\vert 1,0,0 \rangle$ state in the interferometer and by adding a $\psi = \pi/4$ feedback phase on mode $\mathbf{k}_{3}$. Depending on the sign of the derivative, we can discriminate between $\phi_{\mathrm{r}}^{1,(a)}$ and $\phi_{\mathrm{r}}^{1,(b)}$. This leads us to a new probability distribution $P^{(2)}(\phi \vert \{ N_{1}^{(i)}, N_{2}^{(i)}, N_{3}^{(i)} \}_{i=1}^{M_{1}+M_{2}})$, centered in $\phi_{\mathrm{r}}$.
\item[(III)] The last step of the protocol exploits the information acquired in steps (I)-(II). At this stage, our knowledge of the phase is encoded in the distribution $P^{(2)}(\phi \vert \{ N_{1}^{(i)}, N_{2}^{(i)}, N_{3}^{(i)} \}_{i=1}^{M_{1}+M_{2}})$. We then perform the remaining $M_{3} = M - M_{1} - M_{2}$ measurements by adding a feedback phase $\psi$ according to $\mathrm{\phi}_{\mathrm{r}}$ in order to translate the working point, i.e., the total phase $\phi_{\mathrm{t}} = \phi + \psi$ of mode $\mathbf{k}_{3}$, to the value $\phi_{\mathrm{t}} \simeq 2 \pi/3$. This choice corresponds to the maximum of $I_{\phi}^{\mathrm{F}}$ of the system. Finally, by adopting a Bayesian approach we obtain the final distribution $P^{(3)}(\phi \vert \{ N_{1}^{(i)}, N_{2}^{(i)}, N_{3}^{(i)} \}_{i=1}^{M})$ for the phase $\phi$. The value of $\phi_{\mathrm{est}}$ and $\sigma^{2}_{\mathrm{est}}$ is then retrieved according to Eqs. (\ref{eq:bayesian_1}-\ref{eq:bayesian_2}).
\end{itemize}

\subsection{Multiparameter estimation}

Here we briefly present the theory for the multiparameter estimation problem. In this case, the bounds for the estimation of the set of parameters ${\bm \lambda} = (\lambda_{1},\cdots,\lambda_{n})$ are defined by the quantum Fisher information matrix:
\begin{equation}
\mathbf{H}({\bm \lambda})_{\mu \nu} = \mathrm{Tr}\Big[\rho_{\bm \lambda}\frac{L_{\mu} L_{\nu}+L_{\nu} L_{\mu}}{2} \Big].
\end{equation}
Here, $L_{\mu}$ is the symmetric logarithmic derivative with respect to the parameter $\mu$, defined as $\partial_{\mu} \rho_{\bm \lambda}=(L_{\mu} \rho_{\bm \lambda}+\rho_{\bm \lambda} L_{\mu})/2$. The error on the single parameter $\lambda_{\mu}$ is bounded by the quantum Cramer-Rao bound $\delta \lambda_{\mu} \geq [(\mathbf{H}^{-1})_{\mu \mu}/M]^{-1/2}$. When dealing with the simultaneous estimation of the set of parameters ${\bm \lambda}$, the quantum Cramer-Rao inequality on the sum of the variances reads:
\begin{equation}
\sum_{\mu} \mathrm{Var}(\lambda_{\mu}) \geq \frac{\mathrm{Tr}[\mathbf{H}^{-1}]}{M}.
\end{equation}
Differently from the single parameter case this bound is not always achievable. A necessary condition for the attainability of the multiparameter quantum Cramer-Rao inequality is provided by the following identity:
\begin{equation}
\label{eq:weak_commutativity}
\mathrm{Tr}\big[\rho_{\bm \lambda}[L_{\mu},L_{\nu}]\big]=0.
\end{equation}
This condition corresponds to requiring that the optimal measurements for the estimation of the single parameters are compatible observable, which in general may not be satisfied.

Let us now consider the case of $3$-modes and $4$-modes interferometers with two unknown phases $\phi_2$ and $\phi_3$ presented in the main text. For pure state input $\rho_{\bm \lambda} = \vert \psi_{\bm \lambda } \rangle \langle \psi_{\bm \lambda } \vert$ and unitary evolution $\vert \psi_{\bm \lambda} \rangle = \mathcal{U}_{\bm \lambda} \vert \psi_{0} \rangle$ of the form $\mathcal{U}_{\bm \lambda} = e^{-\imath \sum_{\mu} G_{\mu} \lambda}$, where the generators commute $[G_{\mu},G_{\nu}]=0$, the symmetric logarithmic derivative can be evaluated as $L_{\mu} = \mathcal{U}_{\bm \lambda} L_{0 \mu} \mathcal{U}^{\dag}_{\bm \lambda}$, where: $L_{0 \mu} = 2 [(-\imath G_{\mu}) \vert \psi_0 \rangle \langle \psi_0 \vert + \vert \psi_0 \rangle \langle \psi_0 \vert (\imath G_{\mu})]$. In the case of the integrated interferometers, the generators $G_{\mu} = n_{\mu}$ commute, and by directly evaluating Eq. (\ref{eq:weak_commutativity}) we obtain that the necessary condition for the attainability of the multiparameter quantum Cramer-Rao inequality is satisfied.
